# Protection of DVB Systems by Trusted Computing

Nicolai Kuntze
Andreas U. Schmidt
Fraunhofer Institute for Secure
Information Technology SIT
64295 Darmstadt, Germany
andreas.schmidt@sit.fraunhofer.de
nicolai.kuntze@sit.fraunhofer.de

**Abstract**
*We describe a concept to employ Trusted Computing technology to secure Conditional Access Systems (CAS) for DVB. Central is the embedding of a trusted platform module (TPM) into the set-top-box or residential home gateway. Various deployment scenarios exhibit possibilities of charging co-operation with mobile network operators (MNO), or other payment providers.*

## INTRODUCTION

Delivery of digital content to customers is an emerging market with high potential revenues. For various business cases protection of the consumption good as a base for charging processes is required.

This invention report addresses the design of a trusted set-top box which can be reconfigured for various content protection schemes and payment methods. This reconfiguration is done in a trustworthy way. Some implementation variations are presented and discussed.

Digital Video Broadcast (DVB) as the widest spread standard for digital content delivery implements a protection for the digital data. DVB exists in three branches specialized for different broadcasting techniques und formats as satellite (DVB-S), terrestrial (DVB-T), and mobile environment (DVB-H). All branches are equal in its processing of the signal provided. The signal is encrypted by the Common-Scrambling-Algorithm (CSA) which needs a 8 byte seed for initialization (only 6 bytes are used), the so called Code Word (CW). This Code Word is provided by a second algorithm the Conditional-Access-System (CAS). There are many vendors offering CAS Modules for the provider of the content like Cryptoworks or NDS.

CSA was kept as a secret over a couple of years, but was revealed some time ago [1]. Until now CSA is not broken. Microsoft with its DRM solution is a second example for existing systems protecting digital content.

Payment solutions are the aim for the smart card – set-top box combination. This market is dominated by smart card subscriptions. The customer registers the card after purchase at the provider and is able to descramble the digital stream for a certain time. On this basis, pay per view schemes are in use clearing the program for a certain time, e.g. for one movie. The accounting is for instance solved by using value added telephone services. A second way of charging for DVB content is using mobile payment solutions. One (German) peculiarity is the use of debit or credit cards in combination with a feedback channel of the set-top box [2].

In the established DVB scheme, the Conditional Access System (CAS) is assigned to perform this essential role of bridging between the encrypted, digital data stream and a smart card providing the required keys. Due to various different CAS systems the customer needs different cards often with different Conditional Access Modules (CAM). We propose a concept based on trusted computing, to improve on this state of affairs.

## CONCEPTS AND DEPLOYMENT SCENARIOS

TC as it is currently being standardized in the PC and mobile domain offers a highly flexible and reliable security infrastructure enabling various kinds of authentication, authorization, and audit schemes. A Trusted Platform Module (TPM) acts as a hardware anchor and provides a unique identity of the particular device. This anchor provides cryptographic abilities for creation, storage, and usage of asymmetric key pairs. On top of this TPM abilities TC offers a metric to measure the system state and to report this measurement to a third entity. A further main ability is to bind certain data cryptographically to the particular platform and its state. An overview on Trusted Computing functionality can be found in the appendix. Based on these three basic features a *virtualization* of the CAS is feasible. Implementing in this way a *trusted set-top box* in the context of DVB can be done in many variations. An elementary implementation stores the functionality of the CAM and of the smart card as software protected by means of the TPM. If the user requests an encrypted channel, the set top box uses the CAM/smart card software to create the respective code words required by the CAS.

**Online CAS**
A system which emulates the actual CAS systems in software has to enforce that keys used to generate the CWs required by the CAS are kept secret. The security of the proposed system has to be guaranteed even if the algorithm is published. Beide this technical requirement online verification of the access rights as a second requirement is the premise for dynamically granted content access. This leads to an authentication scheme where the CAS asks for permission before the CWs are created. Therefore a connection is established by a communication module granting access to a network.

A protocol solving this problem has to perform the following steps. In preparation an appropriate roll out of the set-top boxes has to happen. During the roll-out a *take ownership* by the provider is required. A network operator can be used instead of the provider assigning user identities and, optionally, performing charging processes. The proposed scenario consists of four parties: customer, provider, charging provider, and network operator as depicted in Figure 1. The network operator issues the set-top boxes to the customers in the same way as they offer mobile devices. The bonding between customer and device is based e.g. on a SIM-card. An online take ownership is as well possible as described in the subsequent section.

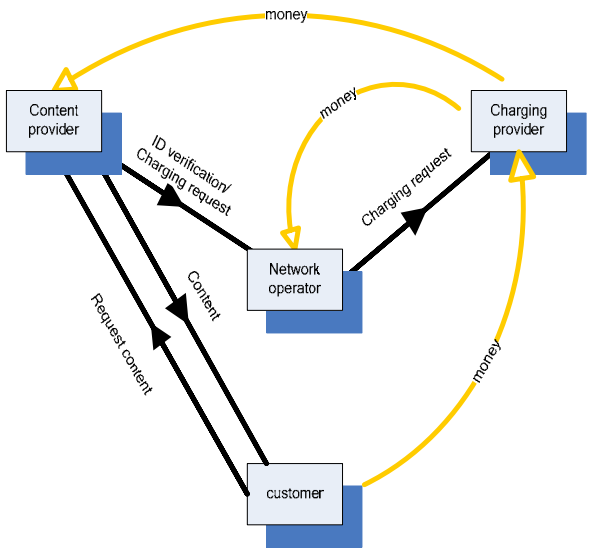

**Figure 1. Online CAS scenario**

The network operator establishes his trust in the device by a network operator issued credential which is produced by the box on request. Based on this underlying trust relation the (M)NO can assure the identity of the set-top box to a supplier. This second (transitive) relation [3] is based on a second credential issued either by the network operator or the trusted set-top box. In either case the network operator in his role as an identity provider (ID) signs this credential which therefore is stored in the trusted set-top box. If a customer decides to consume a certain service it offers this credential to the vendor (V), for instance the content provider. V uses this credential to verify the identity against ID. V then delivers the content and requests charging by ID. In this scenario the user is unknown to V as the credential is only validated by ID. ID does not need to reveal the identity to V. AIKs from the context of the trusted computing standard can be used as such credentials identifying a certain user.

Delivery in this context means that V transfers a secret into the trusted set-top box and adds this secret to the list of accepted credentials as this is known by the actual process of conventional CASs. This secret is sealed in the set-top box by using the sealing functionality of the TPM. This means the issued secret can only be used in the same trustworthy state of the box. The root of trust in this case is the ability of attestation of the integrity to a third party e.g. V or ID.

In an online CAS a tight interweaving between *content access control* and *payment* scenarios is possible. This is discussed in the following section.

**Online take ownership / online registration**
The aim of an online take ownership procedure is to establish a user identifying credential into a trusted set-top box without the need to issue the box over a special infrastructure provided by ID. The customer should be able to buy such a box everywhere she wants. During production every box gets an identifier in form of the unique platform certificate. Based on this initial credential a protocol can be performed to establish a user related certificate. This user certificate is used to identify the user at V. The used protocol establishing this user credential depends if there is a direct or indirect communication between V and ID.

The user certificate can be created after the take ownership process of a trusted platform which binds a TPM to a certain user using a 160 bit authentication value (TPM owner authorization). AIKs are available after the take ownership to be used as credentials testifying the identity of a user. An AIK can only be created offering a valid TPM owner authorization and is a private/public key pair. The private portion is shielded inside the TPM. After this, a privacy CA issues a certificate to assert the relation between AIK and TPM. See the Appendix. For this initial AIK and certificate creation process an online connection to a privacy CA is required. The pertinent protocol has to protect the origin of the key so that it is impossible to fake a TPM.

In the case of a direct communication the system is equipped with a communication device enabling the direct contact to the privacy CA. In this case the mentioned protocols can be used without any restrictions. If the system is not equipped with such a communication device at least a short range communication is required enabling a take ownership over a secondary communication device (SD) like a cell phone. The SD forwards the communication in the respective direction.

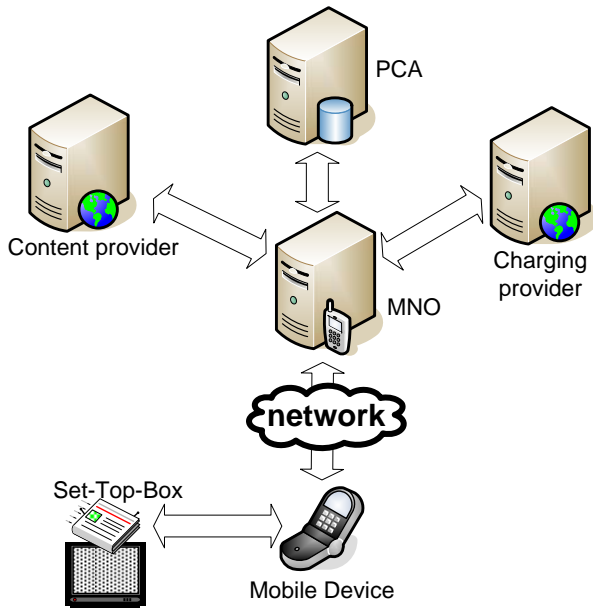

Figure 2. Offline take ownership.

It is important to mention that there has to be a trust relation between the privacy CA and the content providers. In this use case AIKs are used as tickets which enable accounting. Therefore it could be possible that the privacy CA should be able to reveal the identity of a certain AIK, e.g. in case of suspected fraud.

After the take ownership an online registration at the respective provider is required to sign up to a certain service. In this process two goals have to be achieved. First, the identity of the mobile device (and therefore the identity of the user) has to be registered at the service provider. By issuing the AIK and the belonging certificate the identity can be proved and by performing a handshake protocol between service provider and mobile device the origin is testified. The second aim is to negotiate about the details of the subscription. One part of these details is the payment information. The service provider transmits a data structure which describes the available charging models and services. The user selects from this offer, signs the selection with the private portion of the AIK and transmits this to the service provider.

A proof of authenticity of the service provider is also required. Hence it is necessary to sign messages issued by the service provider. A verification of the authenticity can be achieved using known PKI structures or built-in root certificates. Alternatively, the MNO can vouch for the authenticity of a certain service provider replacing conventional PKI systems.

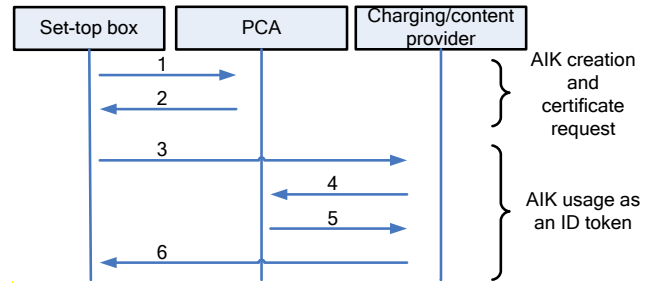

Figure 3. AIK authentication for content access.

Figure 3 shows a basic scheme using AIKs as authentication tokens. The protocol is divided in two phases. Phase 1 is concerned wit the AIK creation and certification, phase 2 is the usage of this token. The set-top box transmits in 1) certain credentials of the platform and the public portion of the AIK to the PCA which verifies the offered credentials and then retransmits (2) an certificate stating that this AIK belongs to a an accepted platform. If the MNO works as PCA the AIK can later be used as a payment credential as the MNO can reveal the identity of the users based on the certificate. This feature is used in phase 2 where the set-top box offers the AIK and the corresponding certificate to a service provider (3). This service provider performs an attestation of the box in this step and then request from the CA the validity of the offered certificate (4,5). 4 and 5 can be performed e.g. using OCSP responder known from PKI infrastructures. Step 6 returns the acceptance information of the service provider (this answer should be signed by him). To validate the signature on side of the set-top box an appropriate certificate must be available in the device. It can be necessary to request this root certificate from a trustworthy third party. After this the set-top box has been successfully registered at a certain service provider.

**Payment Scenarios**

Trusted computing offers the possibility for client-side payment systems. The main idea is that the set-top box handles charging in equivalent to prepaid or pay-as-you-go scenarios. The box meters the consumption of the digital content. Either an internal deposit is decremented or the consumption data is stored and after a certain time read by V and charged. The second alternative requires a trusted time source in the device to enable time stamps for the recorded metering data. This time source can be implemented as an online time authority which offers a time-stamping service. Another option is an internal time source based on the trusted platform. This shall be discussed elsewhere. A third usage scenario is the well known distribution of keys which are descrambling a channel for a certain time.

The first alternative requires a central authority to increment the internal deposit. Therefore it is required to transmit the increment value to the respective box. The increment value is signed by the concerned charging service and encrypted by the public portion of the AIK. These two steps are ensuring that this packet is bound to one device. Preventing a replay attack requires to include (i) to perform an attestation of the particular device by the charging service testifying the integrity of the device and (ii) to integrate a Nonce issued by the charging service to prevent that the packet is used twice. The box stores these Nonces and rejects replayed packets. A basic protocol works as follows. In step (1) the box requests for an increment of the internal deposit, offering the AIK and a corresponding certificate. Step (2) performs an attestation and within this step a handshake testifying the correlation of the AIK and its certificate. At some point the in these first two steps the charging provider has to decide if he accepts this AIK based on a contractual relation to the user or its charging provider and PCA. After step 2 the charging provider has to decide which amount is transferred to the box. This can be done based on a pre-paid scenario, a pay-as-you-go scheme, or other contractual scenarios like charging the telephone bill of the user if the PCA or the charging provider is operated by the MNO.

Storing the consumption in the device is the second alternative. Here it is possible to offer the user the possibility first to consume the content and afterwards to pay for it as it is known from e.g. cellular phones. The stored consumption data are transferred to the charging provider either by a push system where the box is in charge for transferring the data to the provider or by a pull system where the provider requests the data from every box. The most significant difference between these two schemes is the signaling of the box as the pull solution requires a built in communication device like a GSM module. Building on a push solution another device like a mobile phone can be utilized to transfer the data to the charging provider. Important to note in the second case is that the logged data require a special protection against forgery. Therefore each consumption log record needs to be time stamped by a trustworthy time source. After the transfer of the consumption log the charging provider charges the consumer belonging to the system. This relation between consumption data and user is expressed by signing the data with the respective AIK which was first registered at the charging provider (see last section). As a privacy protection after the signing the data can be encrypted by a public key belonging to the charging provider. By this an MNO or any third party listening to the data stream cannot create usage profiles of the customers.

A third alternative is based on usage credentials which are basically descrambling keys inserted e.g. in the conventional CAS implemented in DVB and enhanced by a constraint of e.g. its usage period. Other constraints are also possible. Examples are maximum boundaries of daily usage, or time of usage. Issuing such a restricted key to a system requires (i) an attestation of the system before the key is transmitted to proof the integrity of the system, (ii) a PKI structure to enable the system to verify the origin of the data, and (iii) a charging infrastructure which maintains the keys. Based on the CAS system it can be necessary to store the key in a sealed environment as it is offered by the TPM sealing functions. A variant of this kind of content protection is exhibited in [4]

**ADVANCED FUNCTIONALITIES AND BENEFITS**

A trusted set-top box enables the creation of a universal decoder for broad range of different scrambling systems without the need for the user to physically change the smart card. The virtualization of the CAS hardware is also interesting for reducing the costs of a single set top box. This can sensibly be combined with either a long range communication module, e.g. GSM or UMTS, or a near range communication module like Bluetooth or NFC. This enhancement enables complex scenarios with respect to accounting and charging, some of which are presented in the following list.

- As mentioned, the communication between the trusted set-top box can be performed by an integrated long range communication device or a near range device. Using GSM/UMTS in a box enables the cooperation between a mobile network operator who can offer accounting, charging, and maintenance for the customers and box vendor/DVB provider. The MNO is also able to offer a well tested authentication system in form of the SIM-card known in the mobile domain. Based on this primary identifier other authentication schemes using the TPM are possible as they are introduced in [5].
- Online payment using e.g. credit cards in the set-top box is one of the most important possibilities which are enabled by trusted set top boxes.
- Implementing an online CAS Module which replaces the CAS-Smartcard complex can eliminate the possibility of using fake smart cards as the box can control the access rights by an access control list at the side of the provider. TC provides in this scenario the authenticity and identity of the box and the attestation of its integrity (by the TC methods of remote, or direct anonymous attestation). With these basic assumptions it is possible to implement a trustworthy client side CAS.

- As the set-top box can be considered as trustworthy, such a box enables easy-to-use payment schemes which have the potential to replace existing Pay-per-View schemes. By executing charging on the client side, price schedules can depend on the consumed service or content which can be metered by the trustworthy box. It would as well be possible to recognize debit and credit cards or to prepay various contents.
- Online Algorithm updates for video decryption are an extension of the variants above, which enables on demand reconfiguration for unknown decryption schemes or the exchange of existing ones. The vendor or broadcaster can issue in a standardized configuration file the new algorithm which is to be used in the box to decrypt the incoming video stream. This also enables over the air enhancements and updates of the trusted set top box e.g. faster implementations.

Near range communication devices like Bluetooth require SDs like mobile phone as communication intermediaries to the broadcaster. As it may be cheaper than a dedicated GSM module in each set-top box this could be the better choice. This solution requires charging in the set-top box as described above. Charging and updates are performed by using SD. After this step the set-top box works autonomous.

**Algorithm updates**
As stated in the introduction the scrambling of the digital content relies on the security of the CSA. This algorithm has to fulfill two requirements. The first is to provide a secure scrambling of the data. The second is the real-time environment. The digital stream has to be descrambled in the moment it arrives. This leads to an algorithm the strength of which is not as high as possible and the need to replace the algorithm can be foreseen [5, 6], entailing a complete change of the installed infrastructure on side of the customers as well – a very costly and inconvenient effort.

From the viewpoint of customer satisfaction it is advantageous if it is possible to update a certain device to new required standards and protocols. As introduced in the online CAS section, it is possible to implement the complete CAS as a software module protected by the underlying TC architecture. Depending on the implementation modules of the DVB system can be updated by a software protocol. The update process can be started by the system based on an explicit user request. First, the device transmits a nonce and a device description data block (DDDB) to the update service which is signed by the AIK. Depending on the DDDB the update service creates a data block containing the new firmware of the device and signs it with the signing key of the update service. This data is then encrypted with the AIK and send to the device. The device then decrypts and checks the received data and updates the system. If reprogrammable hardware such like FPGAs the update results in a reprogramming of the hardware. It is important to consider reliable checks of the content of such hardware and hardware based methods with are protecting the content of such hardware against fraud.

The presented concept enables a full modular system where it is possible to replace every part of the descrambling in a trustworthy way. Using for instance Field Programmable Gate Arrays (FPGA) technology in combination with Trusted Computing enables the replacement of every part of the existing DVB architecture. This has the advantage that it is possible to implement algorithms in hardware gaining a speed-up compared with software based implementations. FPGAs are programmed (reconfigured) before they can perform the desired task. Trustworthy implementations have to verify the content of the FPGA before data are transmitted between FPGA and the CAM.

**Multi-CAS management**
New accounting and charging systems can be established using the TPM and the trusted environment. Here the sealing functionality is highlighted, as a replacement (or new specification) for the CAS. One box can house various CAS concurrently *without any extra hardware costs*. Charging can be transferred into the Box by adding specialized software and appropriate hardware. The trustworthy state of both can be proved in the trusted boot process and attested to an external verifier by remote or anonymous attestation. The box can handle the charging process and control the access to the charged commodity. By this, new payment infrastructures can be established, e.g., for home shopping.

**Mobile co-operation**
The presented concepts offer a high degree of co-operation of DVB service providers with mobile networks and their MNOs. This co-operation brings major benefits, though the technology DVB is not common in the mobile domain. The MNO can provide primary services to DVB operaters or broadcaster:
- Management of subscribers
- Potential merging of the subscriber bases
- Charging
- Dynamical control of security features
- Enhanced CAS security
- Marketing co-operation

Furthermore the concept can complement current or development mobile content protection like OMA. In effect it enables MNOs to enter a mobile video broadcasting partial market. The underlying concept is not limited to DVB. It can enhance or substitute existing CAS systems for streaming multimedia applications. It is also possible to combine this with a bonding of the set-top box to a special mobile phone or network operator. A network operator can offer set-top boxes as part of their customer retention.

**Client-side watermarking**

As an extension the box can apply certain digital rights management (DRM) functionalities upon the digital content. These restrictions can be applied by the box dependent on the contract between the partners. These restrictions can for example regulate the count of allowed copies. Depending on the rights purchased by the user different usages of the content (multiple viewing, multiple private copies, quality discrimination, commercial-free versions, etc) can be permitted by application of DRM techniques. As every customer can buy different rights for one digital asset this restriction have to be added on customer side. It could also be intended, as an alternative to 'hard' DRM, to tag a digital asset and by this to suppress the distribution. This leads to the need to include a marking identifying the individual customer, while still preserving his/her privacy. This service can be a major reason to apply watermarking at client-side. This can be implemented with ease within the trusted set-top-box based on our concept.

## CONCLUSIONS

We have shown that turning set-top boxes into trusted platforms according to TCG specifications offers attractive novel possibilities for DVB content distribution, CAS operation, and payment. Prerequisites for implementation seem rather low as todays set-top boxes are already full-fledged PCs and TC will be ubiquitous in the near future.

In a more evolved scenario the TPM could be an integral element of the code word scheme replacing the smartcard.

## APPENDIX

**Trusted Computing Essentials**

Trusted Computing uses a hardware anchor as a root of trust and is now entering the mobile domain with the aim to provide a standardised security infrastructure. Trust as defined by the Trusted Computing Group (TCG) means that an entity always behaves in the expected manner for the intended purpose. The trust anchor, called Trusted Platform Module (TPM), offers various functions related to security. Each TPM is bound to a certain environment and together they form a trusted platform (TP) from which the TPM cannot be removed. Through the TPM the TP gains a cryptographic engine and a protected storage. Each physical instantiation of a TPM has a unique identity by an Endorsement Key (EK) which is created at manufacture time. This key is used as a base for secure transactions as the Endorsement Key Credential (EKC) asserts that the holder of the private portion of the EK is a TPM conforming to the TCG specification. The EKC is issued as well at production time and the private part of the key pair does not leave the TPM. There are other credentials specified by the TCG which are stating the conformance of the TPM and the platform for instance the so called platform credential. Before a TPM can be used a take ownership procedure must be performed in which the usage of the TPM is bound to a certain user. The following technical details are taken from [7].

The TPM is equipped with a physical random number generator, and a key generation component which creates RSA key pairs. The key generator is designed as a protected capability, and the created private keys are kept in a *shielded capability* (a protected storage space *inside* the TPM). The shielded capabilities protect internal data structures by controlling their use. Three of them are essential for applications.

First, *key creation and management*, second the ability to create a *trust measurement* which can be used to assert a certain state toward a, remote party, and finally *sealing* methods to protect arbitrary data by *binding* it (in TCG nomenclature) to TP states and TPM keys.

For the TPM to issue an assertion about the system state, two *attestation* protocols are available. As the uniqueness of every TPM leads to privacy concerns, they provide pseudonymity, resp., anonymity. Both protocols rest on Attestation Identity Keys (AIKs) which are placeholders for the EK. An AIK is a 1024 bit RSA key whose private portion is sealed inside the TPM. The simpler protocol Remote Attestation (RA) offers pseudonymity employing a trusted third party, the Privacy CA (PCA), which issues a credential stating that the respective AIK is generated by a sound TPM within a valid platform. The system state is measured by a reporting process with the TPM as central reporting authority receiving measurement values and calculating a unique representation of the state using hash values. For this, the TPM has several Platform Configuration Registers (PCR). Beginning with the system boot each component reports a measurement value, e.g., a hash value over the BIOS, to the TPM and stores it in a log file. During RA the communication partner acting as verifier receives this log file and the corresponding PCR value. The verifier can then decide if the device is in a configuration which is trustworthy from his perspective. Apart from RA, the TCG has defined Direct Anonymous Attestation. This involved protocol is based on a zero knowledge proof but due to certain constraints of the hardware it is not implemented in current TPMs.

AIKs are crucial for applications since they can not only be used, according to TCG standards, to attest the origin and authenticity of a trust measurement, but also to authenticate other keys generated by the TPM. Before an AIK can testify the authenticity of any data, a PCA has to issue a credential for it. This credential together with the AIK can therefore be used as an identity for this platform. The protocol for issuing this credential consists in three basic steps.

First, the TPM generates an RSA key pair by performing the **TPM_MakeIdentity** command. The resulting public key together with certain credentials identifying the platform is then transferred to the PCA. Second, the PCA verifies the correctness of the produced credentials and the AIK signature. If they are valid the PCA creates the AIK credential which contains an identity label, the AIK public key, and information about the TPM and the platform. A special structure containing the AIK credential is created which is used in step three to activate the AIK by executing the **TPM_ActivateIdentity** command. So far, the TCG-specified protocol is not completely secure, since between steps two and three, some kind of handshake between PCA and platform is missing. The existing protocol could sensibly be enhanced by a challenge/response part to verify the link between the credentials offered in step one and used in step two, and the issuing TPM. The remote attestation process is shown in Figure 4.

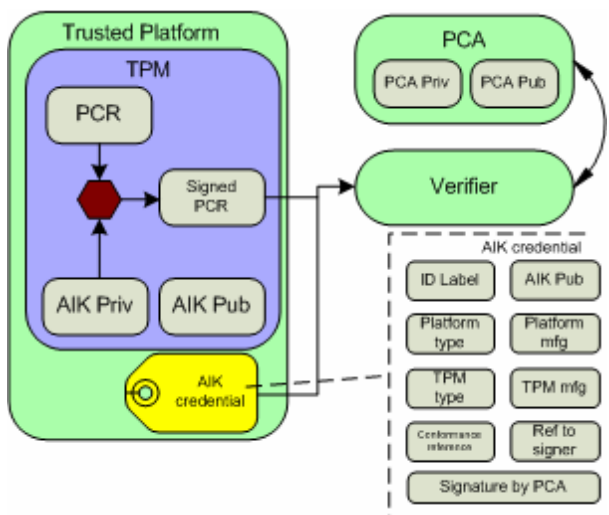

**Figure 4. Remote attestation process.**

Beside the attestation methods TC offers a concept to bind data blobs to a single instantiation and state of a TPM. The **TPM_unbind** operation takes the data blob that is the result of a **Tspi_Data_Bind** command and decrypts it for export to the user. The caller must authorise the use of the key to decrypt the incoming blob. In consequence this data blob is only accessible if the platform is in the namely state which is associated with the respective PCR value.

A mobile version of the TPM is currently being defined by the TCG's Mobile Phone Working Group [8]. This Mobile Trusted Module (MTM) differs significantly from the TPM of the PC world and is in fact more powerful in some respects. In particular, it contains a built-in verifier for attestation requests, substituting partly for an external PCA. Both TPM and MTM are a basis for application architectures. Trusted Computing affects the world of networked PCs but also heavily impacts the mobile industry.